\documentclass[12pt]{article}
\input{epsf}
\usepackage{amsmath}
\usepackage{amssymb}
\begin{document} 
\title{Antiquity, Dark Ages and Renaissance of Light-Front Higher Spin Theory}       

\author{Anders K. H. Bengtsson\footnote{e-mail: akhbengtsson@protonmail.com}}

\maketitle

\begin{abstract}
These notes contain -- apart from some physics -- scattered reminiscences of everyday life at the Institute for Theoretical Physics in G\"oteborg in the early eighties. The text has been in the making for many years. Some of it was written inspired by the Lars Brink 70-fest at The Solvay Institute in Brussels in 2014, but was never finished or published. Parts of the old text has now been replaced by a review of the quite unexpected recent renaissance for light-front higher spin theory. Lars sadly did not live to experience more than the very beginnings of this quite remarkable development. The text was finalized in the spring of 2023.
\end{abstract}

\pagebreak

\section*{Fortunately}\label{sec:Fortunately}
I was very fortunate to become one of Lars Brink's graduate students during the early 1980's. It was a very different time from now. As I remember it, I just strolled into his office. 

No, that's not really true. It was actually preceded by quite some vacillating on my side and an ultimatum from one of the towering professors of the Institute at that time. This must have been sometime in the late spring of 1979. Lars sat in his big office at the end of the infinite corridor on floor 8 of the Origo building, shoes in an open drawer, and told me about the flow of the coupling constants and unification of the forces somewhere below the Planck scale. They all seemed to converge at a common point. This was a time when the Standard Model was just a few years old and it was all new and brightly shining. And the future looked even brighter with supersymmetry recently theoretically discovered and looking promising.

And in the fall of that year, Weinberg, Glashow and Salam received the Nobel Prize for the electroweak unification theory. They gave public talks at Chalmers and visited the Institute for Theoretical Physics and Mechanics (ITP for short) and we were invited by Abdus Salam to the Mosque in V{\"a}stra Fr{\"o}lunda. It was a very privileged time and a very inspiring environment. G{\"o}teborg is not exactly the center of the world but it felt like it wasn't that far off those years. Nobel Prize winners passed by.

So how did I end up in Lars' office? This is my story. When I was about five years old my grandmother started to bring home old radio sets and telephones for me to take apart and try to put together again. She had a large cupboard full of these things. I don't know from where she got that unusual idea, but I like to think that it gave me a scientific outlook at an early age. I remember when we had the first physics classes in school and learned about the Archimedean principle, the rule of levers, block and tackles and stuff like that. I couldn't understand how anyone could find such things difficult or hard to understand. I found it all self evident. Obviously reality couldn't be organized in any other way. I told this to the teacher and he looked at me disbelievingly. Perhaps it was because I was playing with Meccano and other model building kits. Around this time a friend asked me what I wanted to be when I grew up. I had just read a comic book where the hero said that he was a ''nuclear physicist'' -- so that was my answer too. After that there was of course no turning back. A few years later I read all popular science books about physics and mathematics that I could find at the local library. There were lots of them, most of them written in the fifties and sixties and then appearing in Swedish translations. 

The ultimatum that I had to act on was the following. I had become --  as many other young people at that time -- involved in the Swedish environmental movement. During my last year at Chalmers University of Technology I learned about a research group in Physical Resource Theory that sounded interesting (the term sustainable development wasn't coined yet, but that was part of what it was about). The group was headed by Karl-Erik Eriksson who was really a professor in theoretical physics. To make a long story short -- of which I don't know very much anyway -- other professors at the institute were not so happy with this state of affairs. Into this controversy I now stepped unknowingly. I did my undergraduate project on energy forecasting with Karl-Erik. I had many conversations with him and his students, but I couldn't make up my mind to join their group as a research student. There was something vague about it, which was perhaps not so surprising for a new field, and my interest in theoretical physics had awakened again. One day I was summoned to Stig Lundqvist's office. He simply told me to make up my mind -- and do it fast -- and decide which group I wanted to join. So I did. It may have been the very same day that I walked down the corridor and talked to Lars. It was the best choice. 

The fact that Karl-Erik was an elementary particle physicist also helped turn the tables. He said something to the effect that ''There's nothing wrong with getting a real education.''. Another person instrumental in keeping me on-track was Arne Kilhberg. Around this time I had applied for a job in Stockholm at a government board for energy (Energiverket) and got it. I went to Arne for advice and after that turned down the offer. But the amazing fact remains -- all this hesitation on my part notwithstanding -- Lars accepted me as a student right away. Thanks for that!

A funny thing is that for a few years in the gymnasium I actually lived across the street from Chalmers and the ITP. I had no idea that just a few minutes walk away, people spent their days working on the same kinds of problems as Dirac, Schr{\"o}dinger, Heisenberg and Pauli had -- the people in the popular science books that I've read. I had thought that quantum mechanics and relativity was only studied at places like G{\"o}ttingen, Cambridge and Princeton. It was beyond imagination that it could be done in G{\"o}teborg.

I did not think about it consciously at the time, but then in the early 80's there was a strong connection back to the heydays of particle physics in the 50's and 60's. Professor Jan Nilsson was of that generation and sort of carried the air of those distant times around him, having worked on weak interaction theory. I think one of the things that drew me back to theoretical physics was a course he held in Elementary Particle Physics that I took in my last year as an Engineering Physics student at Chalmers. His lecture notes -- hand written -- were very beautiful and an exemplar of clarity as were the lectures he gave. The Dirac equation never looked so beautiful as in Jan's handwriting. But sometimes the lectures were cut short by a secretary coming in and telling him there was a phone call from China. He rushed off and did not come back more that day. 

\section*{The mass-shell}\label{sec:TheMass-shell}
I shared a big office with Olof Lindgren who was a few years my senior and contemporary with Bengt Nilsson. I'm not sure if it was the biggest office in the department, but I certainly had the biggest office desktop. When I, after a few months moved up from the seventh to the eight floor, one of the secretaries asked me -- she did look a bit guilty about it -- if it was OK if I could do with an awfully large desk that no-one else apparently wanted. Even then I must have seen the opportunities although I hadn't yet done any serious calculations. Sitting at it I couldn't reach the far ends with my arms stretched out. As we all know, doing theoretical physics requires time and space. Time, in particular if you are a slow thinker (as I am -- sad to admit) and space to spread out the sheets of calculation on. Later, when teaching and students complained about calculations running to more than a few lines, I could tell them about calculations running to hundreds of pages and requiring the desktop of a president.

I don't think I was in any way representative of Lars graduate students around that time. We seemed to be a quite disparate bunch of guys, and one gal, Anna Tollsten. Or perhaps we were all a bit odd -- we who choose to try our hands and wits on high energy physics. Why would any sane person do such a thing? Well, out of curiosity of course -- the only sustainable incentive for research. 

I bonded well with Olof.  I liked his laid-back attitude. Theoretical physics is a very competitive subject and the competition is not always friendly. You're supposed to be smart, and many are of course, but sometimes people were flaunting their smartness. I didn't like that. Although I wouldn't say it was so with Lars' students. He seemed to attract young people who were both independent and generous, in particular so when measured relative our young age. 

And Olof couldn't care less -- he saw through all such silliness. I even liked his habit of whistling (most of the time at least) when calculations went well, which they did now and then. When the whistling did get annoying I took up whistling myself, but Olof  didn't seem to get the clue. By the way, when I met Olof some years ago at an institute reunion and we talked about old times, I understood that he saw me just the same way as I saw him. Funny isn't it?

And then there was the sofa in our office just to the left inside the door. I gave it the name {\it the mass-shell}. We took turns sleeping on it after lunch, the code phrase being: ``I'm on the mass-shell.''.

But it wasn't very comfortable. It was a two-seater and both Olof and I are quite tall. We were soon off-shell again.

We were many graduate students at that time. I may not remember all (there is certainly one name I cannot recollect, I think he left before earning his PhD) but including Robert Marnelius' students there was Bengt Nilsson, Olof Lindgren, Ingemar Bengtsson, Stephen Hwang, Lars Johansson, Mats {\"O}gren, Martin Cederwall, Anna Tollsten, Ragnheidur Gudmunsdottir, Theresa Vallon (who was a post-doc). Ingemar pointed out to me that it was considered unusual in those days with three women in the group. It is not so unusual any longer, fortunately. To add to the mix, two professional philosophers joined the group, for second PhD studies, targeted at the philosophy of quantum mechanics, but Ingemar Bengtsson has written more about this in his text.

But I did not socialize as much as I perhaps should have done. Being a native to the town and having my family and other friends and the engagement in the environmental movement, the incentive was not that strong. So most likely I missed much of the stuff that make for good gossip.

Olof graduated and left for Israel as a post-doc. Martin Cederwall moved in and took up his position by the window. We went along well. He did not whistle at all so I stopped doing it. I did not want to disturb him in the huge calculations he undertook seemingly without effort. The winter 83-84 was cold and the heating in the office did not work that well. I remember Martin working with gloves on -- fingertips sticking out. I stayed at home all of January and February writing up my thesis.

Bo Sundborg came later, I met him when I came home from London. We had discussions on physical principles behind higher spin theory. Funnily enough, we continued these discussions in Firenze in 2013 at a higher spin workshop, as if not twenty-five years had passed. 

But then in 1988, the group as I knew it had broken up as people had graduated and dispersed. The era between the '79 Nobel Prize on electroweak unification and '84 Nobel Prize on the discovery of the W and Z bosons had come to an end, as all eras do. I may be completely wrong, but somehow what was called the first superstring revolution in 1984 had upped the ante, and some of the playfulness seemed to have disappeared. The race was on for the superstring unification Nobel Prize. Stumbling along at the rear of that exodus wasn't a tempting prospect for me. Fortunately, a different road to travel had opened up during my graduate studies.

\section*{Higher spin in early ``Antiquity''}\label{sec:HigherSpinAntiquity}
I was very lucky because without me knowing it, another young man with the same surname, had had a similar conversation with Lars. Ingemar Bengtsson and I became friendly competitive collaborators. Of us, Ingemar was -- and still is, I believe -- the deep thinker. Upon reading a draft of this text, Ingemar wrote back to me that he remember me as the one who had ideas about concrete things (i.e. calculations) to do. So it seems I brought something to the collaboration. 

I learned something very good from Ingemar: having the patience of digging out and reading old references. I think we made this into an art and I still try to practice it. One reason is that it is fun. It is interesting to read old papers that may contain seeds of ideas that come to fruition long later. 
Referring to old papers lends your own papers an air of erudition and generosity -- even some good-humored eccentricity. Only referring to what's been on the archive the last year or so, may make your papers look the opposite: uneducated, ungenerous and conventional. And I think it is good for science on a general level. Research is not just about getting new results -- it is also about refining and understanding old results in new contexts. There is, or should be, a humanistic aspect to all of science.

We came into contact with the higher spin problem in the fall of 1982. Ingemar supplied me with pieces of memory that I had forgotten. Olof and Ingemar and I were in Trieste in September '82 and Peter van Nieuwenhuizen said something in a talk about spin $5/2$ and that ''non-covariant methods'' was the only hope. All three of us thought about the light-cone, according to Ingemar's recollections. I can vaguely remember it. What I do remember are the buckets that were moved around to catch the dripping water from the leaking roof of the lecture hall.

Light-front methods were employed at that time in the superstring theory of Green, Schwarz and Brink, and Lars was certainly a strong proponent of the method. The strength of the method came to the fore with the 1983 proofs of the ultraviolet finiteness of $N=4$ super-Yang Mills theory \cite{Mandelstam1983,BrinkLindgrenNilsson1983}, of which Lars produced one of the proofs with Bengt Nilsson and Olof Lindgren. The work was based on the construction of the theory in the light-cone gauge \cite{BrinkLindgrenNilsson1983N4}. It was after this paper that Lars decided that we should ``do everything on the light-cone''.   I got my own first problem of finding the cubic supergravity interaction in the light-cone gauge. There was a parameter $\lambda$ in the theory with the interpretation of helicity which I set to the value 2. 

Actually, to be completely honest, for a couple of weeks I happily calculated away with the value 0. Such was the depth of my ignorance. I realized the error during a lunch at Chalmers. I think I must have started to get suspicious when I, as an exercise, worked through super-Yang-Mills with $\lambda=1$. It did indeed seem a bit odd to have zero for supergravity. I asked Robert Marnelius what is the helicity of the graviton and he of course answered 2. Back at the desk I started to rework the whole thing. I never told anyone about this embarrassing mistake -- no need to take any risks -- I might have been thrown out! 

There was, however, a reason why I did not spot the error right away. In the complex formulation that we used, the two transverse derivatives were represented by a complex conjugated pair $(\partial,\bar\partial)$. My cubic supergravity vertex had two transverse derivatives in it -- just as it should for a gravity theory. But they came in the combination $\partial\bar\partial$, and the angular momenta of these two derivatives are $+1$ and $-1$ respectively, adding to zero angular momenta for $\partial\bar\partial$. With the correct value for the helicity $\pm2$, the vertex comes with the combinations $\partial\partial$ and $\bar\partial\bar\partial$ distributed over the fields.

But it is generally a good thing to commit errors because they are what you learn from. This particular error made it obvious to me that $\lambda$ could take any non-negative integer value, and I had a rough idea of what would be the result if that was the case. The number of transverse derivatives in the cubic vertex must equal the helicity. But I did not really understand the significance. Higher spins were discussed during a couple of lunch-time conversations in early 1983 about what to do next. I particularly remember one such occasion -- at a pizza joint on Gibraltargatan just outside the Institute -- where Lars said that we should do higher spin. Ingemar remembers a discussion at Lars' favorite down-town lunch time place: Tai Pak. My comment was ``I can do that.''. However, it was Ingemar -- who understood the real significance -- who started to do the detailed calculations with arbitrary $\lambda$. Having got stuck one late night (with the $J^{i-}$ generator) he called me on the phone and I knew what to do since I had the corresponding supergravity calculation in front of me at my desk at home. This was actually a Saturday night and we convened in  Lars office at the institute to sort things out on Sunday morning. As I remember it, Ingemar had already drafted the paper. There was a sense of urgency. This is the formula for the cubic self-interaction for spin $\lambda$

\begin{equation}\label{eq:CubicInteractionLightFront}
\int d^4x\sum_{n=0}^{\lambda}(-1)^n{\lambda\choose n}(\partial^+)^\lambda\phi\,\Big[{\partial\over\partial^+}\Big]^{(\lambda-n)}{\bar\phi}\,\Big[{\partial\over\partial^+}\Big]^n{\bar\phi}+\mbox{c.c.}
\end{equation} 

We wrote two papers \cite{BBB1983a,BBB1983b} that contained the first positive results on massless higher spin self-interactions. The second one is about supersymmetry extended beyond $N=8$. It turned out that the light-front superfields of $N=4$ and $N=8$ could easily be generalized to $N=12,16,20,\ldots$ where $N=12$ corresponds to spin $3$ as the highest spin of the multiplet. This must have been Ingemar's idea. It was certainly Ingemar who realized that the cubic interactions could be done also in the supersymmetric case and he apparently did it during a Friday afternoon group meeting. I don't remember who gave the talk that day, but I remember Ingemar passing small pieces of paper to Lars and me containing cryptic notes. First came $N=12$. A little later $N=16$. And then $N=20$ et cetera. I must admit I was a bit annoyed at the time. It took away the fun of doing the calculations yourself, but at the same time it seemed amazing that the $N=8$ barrier could be breached so easily. And the result itself is strange. I don't think anyone has followed it up in the higher spin community -- and I still don't know what to do with it.

\section*{Onshell or offshell}\label{sec:OnshellOffshell}
In connection to our work on the light-front we had many discussions about whether we were really on-shell or off-shell. The light-front is a bit puzzling. Conventionally, on-shell means what it says: the four momentum $p_\mu$ satisfies the mass-shell condition $p^2=m^2$ or $p^2=0$ for massless particles. In the light-front frame this constraint can be solved as $p^-=p\bar{p}/p^+$ with complex transverse momentum $p$ in four dimensions. When we worked out the generators of the Poincar{\'e} algebra and checked the closure, $p^-$ was everywhere replaced by $p\bar{p}/p^+$. This can be considered as an Hamiltonian formulation of dynamics with $x^+$ the light-front time \cite{Dirac1949FormsRelDyn}.

But this is for free particles and fields. As soon as interactions are considered the equation $p^-=p\bar{p}/p^+$ doesn't make much sense. In classical field theory it can be interpreted as an Hamiltonian
\begin{equation}\label{eq:LFHamiltonian}
\mathcal{P^-}=\int\varphi\frac{\partial\bar{\partial}}{\partial^+}\bar{\varphi}+\mathrm{interaction\;terms}
\end{equation}
generating field equations through Poisson brackets.

For interacting classical fields, on-shell ought to mean: satisfying the classical field equations following from varying an action, in this case $\mathcal{P^-}$ integrated over light-front time. But classically, a system cannot really be off-shell, the equations of motion are always satisfied. There is of course the freedom of initial and boundary conditions, but apart from that, on-shell is all there is. The action is a clever trick for deriving the equations of motion by claiming that the system follows the path minimizing the action. 

The solution to the equation of motion extremizes the action, but it is in quantum mechanics that the action takes on a deeper significance, something Dirac suspected \cite{Dirac1933a} and Feynman later exploited \cite{Feynman1948PathInt}. In quantum mechanics a system can be off-shell, that is, not satisfying the classical equations of motion. In a light-front perturbative QFT, off-shell means going off the light-front free mass-shell in internal lines, just as in covariant perturbative QFT. More generally, in constrained systems, such as higher spin field theory, on-shell refers to the surface of field configurations obeying the field equations, foliated by the gauge orbits determined by the gauge constraints. In light-front field theory all these constraints are explicitly solved.

What it all comes down to in light-front field theory is that we only work with the physical components of the fields. These components then -- in quantum field theory -- can go off-shell in momentum space.

In four dimensions fields of any spin have just two physical components of helicity $\pm\lambda$. Spin zero is a bit ''odd''. We can have a complex spin zero field, and then all massless fields are two-component. Or we can think of helicity as a variable that ranges over the integers and half-integers, and then there is just one component of each helicity. Is this ''two-componentness'', or ''one-componentness'', in four dimensions just a happenstance fact of the representation theory of the Poincar{\'e} group, or does it tell us something about the dimension of space-time?

\section*{Late Antiquity}\label{sec:LateAntiquity}

A few years later, in 1986 in London, Ingemar and I together with Noah Linden returned to the problem of integer higher spin and wrote a paper \cite{BBL1987} where we listed all possible cubic interactions between massless higher spin bosons in four dimensions -- including mixed $s_1$--$s_2$--$s_3$ interactions. In that paper you also find the result that the spin 2 cubic vertex is the square of the spin 1 cubic vertex. Indeed the spin $s$ vertex is the spin 1 vertex raised to the power $s$.

Ingemar hadn't arrived in London when the work started in the autumn of 1986. It actually started with Noah Linden enrolling me in a technically demanding computation of the string $J^{i-}$ generator. Always thinking about higher spin, I got the idea that the methods Noah taught me could be used for the massless higher spin theory. I first tried a string like vertex which did not work. Noah wasn't very happy with this side project to begin with, but I think I convinced him of its importance.

So what is a ``string-like'' vertex. In the formalism used, there are oscillators $\alpha_r$ and momenta $p_s$ with $r,s$ indices labeling the ``string'' fields $\phi_r$ entering the cubic interaction (space-time indices suppressed). A string-like vertex contains terms of the form $\Delta\sim Y^{rs}\alpha_r\cdot p_s+Z^{rs}\alpha_r\cdot\alpha_s$. The Yang-Mills cubic interaction are then produced by the combination $Z^{rs}Y^{tu}\alpha_r\cdot\alpha_s\alpha_t\cdot p_u$ resulting from expanding $\exp(\Delta)$. A this time, I.G. Koh and S. Ouvry were quick to study a covariant string-like vertex for massless higher spin fields in the paper \cite{KohOuvry1986}. They did however not include any term $Z^{rs}\alpha_r\cdot\alpha_s$ in the vertex. Therefore no Yang-Mills interactions were produced. Now I found that it followed from the Poincar\'e algebra that in the light-front formulation $Z^{rs}=\delta^{rs}$ (which turns out to be the case also covariantly). This confused me, until I, in an act of desperation, tried the vertex operator $\Delta\sim Y^{rstu}\alpha_r\cdot\alpha_s\alpha_t\cdot p_u$. Then it all worked out nicely! Incidentally, expanding $\exp(\Delta)$ then yielded all higher spin cubic self-interactions, and with some completions, all cubic higher spin interactions whatsoever can be reproduced. 

Indeed, since the new vertex produced interactions between different spin $s_1$--$s_2$--$s_3$, Ingemar decided to compute all possibilities. This list we included in an appendix, commenting on the nature of the interactions. It is a strange twist of history that the list includes minimal gravitational interactions, with two derivatives, of higher spin fields of the type $2$--$s$--$s$ for arbitrary spin $s$. 

There are four possible types of interactions \cite{BBL1987}, namely
\begin{subequations}\label{eq:IngemarFormulae}
\begin{equation}\label{eq:IngemarFormulaeScalars}
\int\bar{\phi}_0(p_1)\phi_0(p_2)\phi_0(p_3)+\mathrm{c.c.}
\end{equation}
\begin{equation}\label{eq:IngemarFormulaeYMGRHS}
\int\frac{\gamma_1^{\lambda_1}}{\gamma_2^{\lambda_2}\gamma_3^{\lambda_3}}{\mathbb{\bar{P}}}^{(\lambda_2+\lambda_3-\lambda_1)}\bar{\phi}_{\lambda_1}(p_1)\phi_{\lambda_2}(p_2)\phi_{\lambda_3}(p_3)+\mathrm{c.c.}\;\;\text{with}\;\;\lambda_2+\lambda_3>\lambda_1
\end{equation}
\begin{equation}\label{eq:IngemarFormulaeNonSelfInt}
\int\frac{\gamma_2^{\lambda_2}\gamma_3^{\lambda_3}}{\gamma_1^{\lambda_1}}{\mathbb{\bar{P}}}^{(\lambda_1-\lambda_2-\lambda_3)}\phi_{\lambda_1}(p_1)\bar{\phi}_{\lambda_2}(p_2)\bar{\phi}_{\lambda_3}(p_3)+\mathrm{c.c.}\;\;\text{with}\;\;\lambda_2+\lambda_3<\lambda_1
\end{equation}
\begin{equation}\label{eq:IngemarFormulaeFieldStrengthPower}
\int\frac{1}{\gamma_1^{\lambda_1}\gamma_2^{\lambda_2}\gamma_3^{\lambda_3}}{\mathbb{\bar{P}}}^{(\lambda_1+\lambda_2+\lambda_3)}\phi_{\lambda_1}(p_1)\phi_{\lambda_2}(p_2)\phi_{\lambda_3}(p_3)+\mathrm{c.c.}
\end{equation}
\end{subequations}

Here, $\int$ stands for the momentum integrations and momentum conservation delta functions, and $\phi_\lambda(p)$ for a helicity $\lambda$ field. The complex conjugated field $\bar{\phi}_\lambda(p)$ carries helicity $-\lambda$. We use the convention that $\lambda$ is non-negative. The transverse momenta are represented by $\mathbb{P}$ and $\bar{\mathbb{P}}$, which are combinations of $p$, $\bar{p}$ and $p^+$ that make the kinematical part of the Poincar\'e invariance manifest. The list includes all possibilities for cubic interactions excluding field redefinitions of the free field theory. Such field redefinitions involve powers of $\mathbb{P}\bar{\mathbb{P}}$.

The first type (a) is just scalar $\phi^3$.  The third type (c) is a bit odd, it does not contain any self-interactions. The fourth type (d) corresponds to models obtained by taking powers of field strengths in the covariant formulation. Possible counter-terms can be found here.

The second type (b) is the one that contains Yang-Mills and Einstein three-point couplings and their higher spin generalizations. It also contains all sorts of cubic interactions between different helicities, most notably the gravitational interaction for arbitrary integer spin with two transverse derivatives. This type of interaction was long thought to be impossible in Minkowski space-time due to covariant no-go theorems \cite{AragoneDeser1971,AragoneDeser1980a,AragoneDeser1979a,AragoneDeser1980b}.

We did not point out this fact in particular. The existence of these cubic gravitational interactions remained forgotten for 25 years, until I rediscovered them in 2011 when revisiting the light-front approach. There is a Lars twist to that story. But before that, the dark ages of light-front higher spin set in.

In other contexts (in my second book on higher spin theory, and in a talk at the Higher Spin Gravity online-club), I've tried to explain why we did not stress the occurrence of the gravitational cubic interactions. It may seem inexplicable, but speaking for myself, my focus was elsewhere. The main result of the 1987 paper, as I saw it, was the non-string like nature of the vertex operator $\Delta\sim Y^{rstu}\alpha_r\cdot\alpha_s\alpha_t\cdot p_u$ and the immediate consequences of that. This paved the way for an approach to the covariant vertex through BRST techniques, which was the next goal I set myself \cite{AKHB1988}.

All the same, I think we did one big mistake by not broadcasting our results more, and that includes the 1983 results. The only published occasion where the 1983 results were reviewed -- as far as I know of -- was a paper Lars wrote as a contribution to the Fradkin 60th birthday volume \cite{LarsBrink1987HigherSpin}. Then again, there was nowhere to broadcast. A higher spin community did not yet exist. The time for higher spin lay in the future. Mikhail Vasiliev had to work for almost twenty years before a new generation of young theoreticians took notice. And then the  AdS-approach in itself was a bit disconcerting: did it really mean that Minkowski higher spin was a blind-alley? Towards the end of the eighties, I lost confidence. During the 90's I worked (without publishing) on BRST approaches to AdS singleton theory. I did not manage to do what I wanted to do.

Back to 1983, after having found the light-front vertices our focus shifted to the covariant theory. We tried to do covariant spin 3. Not very much came out of this during our time at the ITP. I studied the covariant spin 3 gauge algebra and managed to show that it cannot close on only spin 3 fields \cite{AKHB1985}. This piece of work together with hints to the same effect in Fronsdal's papers \cite{Fronsdal1979conf}, convinced us that in order to have self-interactions for gauge fields of higher spin, an infinite tower of such fields with ever increasing spin is needed. 

Then the paper by Berends, Burgers and van Dam appeared in 1984 with the covariant cubic spin 3 vertex. The existence of cubic interaction terms for a single spin is not in conflict with the infinite tower of fields, the reason being that the non-closure of the gauge algebra probes the quartic level of interaction.

\section*{The Library at the 8th floor}\label{sec:TheLibrary}
The Internet has its strong sides, but it is not as charming as a real library. One of the best things with the ITP in the 80's was the Library housed on the 8th floor in the Origo Building with a fantastic view over G{\"o}teborg. You could also watch the air-pollution inversions when they occurred. 

The library was run by Giorgio Peresutti. Everything was in neat order and there were lots of books -- all the classics -- and new books arrived orderly. All the important journals were there -- The Physical Review, Reviews of Modern Physics, Nuclear Physics, Physics Letters, Journal of Mathematical Physics, Communications in Mathematical Physics, and many more with volumes running back for decades. All the knowledge of theoretical physics was stored in that little library. 

And then there was the preprint shelf. Where one day disaster struck. In those days, preprints were sent to CERN and SLAC who compiled lists that were distributed around the world by mail weekly. So people could write a special post-card to you and request your preprint, and you could do the same. That was great fun. But you also sent your preprint directly to various institutes. And the ITP got preprints that Giorgio put up on the preprint shelf.

Episodic memory is a strange thing. When first writing this passage, I didn't remember exactly how I got the message about the disaster. But I remember the phrase Ingemar used to describe it: ``Today disaster struck at the preprint shelf.'' Ingemar has reminded me of what happened. 

The disaster actually struck at the CERN preprint shelf. Ingemar was there in the fall of '84 and I got the message through an ordinary letter, that is, a piece of folded paper with hand-writing on it delivered physically in a stamped envelop. Digging into my archives I found the letter Ingemar wrote, beginning with the phrase just quoted. Anyway, both Ingemar and I had been struggling with covariant higher spin (although I don't remember us ever comparing notes). The obvious thing to do was to work on spin 3. In principle it was easy enough. The field has three indices and the cubic interaction should have three derivatives so it was just a matter of making an ansatz for the interaction and for the gauge transformation and require the interaction to be invariant. But there are a lot of terms to choose from and I at least made a too restrictive ansatz. And terms with different derivative structure could ''interfere'' with each other due to partial integrations in the action, and then there was the problem of field redefinitons. We worked by pen and paper -- no computer algebra at that time. And the Dutch group got there first, and their method and analysis took advantage of, now well-known, general aspects of the problem.

The phrase about the disaster at the preprint shelf stuck in my memory, and a few years later something similar happened to me. I was at Queen Mary College as a post-doc and I had just sent off my paper on the BRST approach to covariant higher spin fields to Physics Letters, when at the QMC preprint shelf I picked up a preprint by Ouvry and Stern containing similar results. And they beat me to the press by a few weeks so I have to cite their paper first as a matter of good sportsmanship \cite{OuvryStern1986a,AKHB1986a}. 

In connection to this piece of work I had a ''step onto the bus'' experience. I lived in Kentishtown in North London and my first son Olof was with a Swedish lady up in Parliament Hills during the days when my wife and I worked. I used to walk up Highgate Road to fetch him in the afternoons, sometimes jumping onto a bus. I had been struggling for weeks to find some way to write the Fronsdal actions and field equations in a compact way that naturally included them all. There was a preprint by Thomas Curtright that had the equations for various spin formulated with auxiliary fields that much later became known as triplet systems. There was also a preprint by Pierre Ramond that I pored over. I also knew the basics of BRST and string field theory -- the latter was one of the hottest topics in town at that time with Peter West giving seminars on the latest advances. Suddenly one spring day, stepping onto a bus, I saw it all clear: how the Virasoro generators and algebra could be truncated (by taking a zero-tension limit) to yield a free field theory for all integer spin fields in a neat BRST, Warren Siegel-inspired \cite{Siegel1986a}, $\langle\Phi|Q|\Phi\rangle$ theory. 

That day I did not stop with my son on the way home to watch the trains bound for Kings Cross pass under the railway bridge at Kentishtown station. Later in the night, at the kitchen table, Fronsdal's equations fell out in triplet form. It was a joy to write the paper. But I wrote a speculative section at the end which Michael Green advised me to cut out, which I did.

But I've lost direction. I was writing about the library. I really miss it very much. It was a real luxury to have this good library just around the corner of the corridor from your office. Since all the important journals were there with issues running back a long time it was indeed very natural to dig up old references. No fuss with down-loading or password protection -- in a way it was easier to get hold of old references than through the internet today. Perhaps it was what inspired us to dig up all the old papers. There was simply no excuse for not doing it as they were all there, begging for your attention. When calculations got stuck or became to boring to bear, you could always sneak away to library and read something old and dusty, researched and perhaps forgotten a long time ago.

\section*{Dark ages of light-front higher spin theory}\label{sec:DarkAges}

Thinking about it in retrospect, it is a bit strange that we did not attempt the quartic computation in earnest. The vertex operator formalism should have been a suitable launch pad. Myself, however, I turned to the covariant theory in the BRST formulation. Back home in Sweden, after the post-doc years with Mike Green at QMC, I returned to the light front, but soon lost heart. I got a research assistant position in Stockholm, most likely with Lars pulling strings from behind the scenes. No one was doing higher spin in Stockholm at that time -- nowhere else in Sweden for that matter. I wrote a few papers, and then ran out of ideas. The last paper, on twistors and higher spin, was rejected by a referee in a very rude manner. Such was the times. I did not want to move permanently to Stockholm with my family, so I gave up the position before the first four years were up. Instead I started to work as a high school teacher. This left time for some low intensity research. But I was essentially out of the paper reading and writing circuit.

Then I got a letter from Ingemar, containing a paper by Ruslan Metsaev on quartic interactions for higher spin on the light-front. I read it one Friday afternoon at Alhstr\"om's caf\'e in downtown G\"oteborg on my way home from the high school. Ingemar had scribbled a funny comment in the margin that he could not find the ``answer'' in the paper, ``unless it was in the last reference'' which was a reference to a forthcoming paper (there is a second paper, but it is not clear if it is the one referred to). Neither could I find the answer. I filed the paper away and forgot all about it.

I don't know, but it seems that Metsaev himself forgot about the two papers, because there are no more follow ups. Instead he wrote more papers on various aspects of the cubic interactions: higher dimensions, supersymmetry and massive fields. Apart from this, no one worked on light-front higher spin during the 1990's and up until around 2015. Twenty years of dark ages.

\section*{New light}\label{sec:NewLight}

In the summer of 2011, I went to the Strings 2011 conference that happened to be held in Uppsala. I walked to the Uppsala Concert and Congress Hall in the early summer morning, and not seeing anyone I knew, I just strolled along the indoors balcony surrounding the lecture hall looking out at the panorama over Uppsala. Lars turned up, and we chatted away for a minute or two. Then we were approached by a young physicist, Sung-Soo Kim, who introduced himself as working with Lars on gravity counterterms in the light-front formulation. He started to describe a problem they had with the light-front Poincar\'e algebra. The idea of computing counterterms on the light-front was not new to me. I had looked at it while writing up my thesis, and of course, they were all included in the list Ingemar had computed back in 1987. So with Uppsala at our feet we discussed the problem and Sung-Soo took out our old 1983 paper. Now Lars, apparently wanting to save me from embarrassment, said something to the effect that ``Anders may not remember the details of those old calculations'' meaning the 1983 work. However, I realized that the problem they had was connected to the $J^{i-}$ generator. Indeed, it was the same stumbling block that confused Ingemar a long time ago. The problem is actually quite simple. There are two pieces to $J^{i-}$, a dynamical spin part $S^{i-}$ that you must determine, and therefore put your focus on. However, there is also an orbital dynamical part that is $x^iH$ in terms of the Hamiltonian $H$. For some reason, it is easy to overlook the presence of this piece of the dynamical Lorentz generators. 

This conversation lead to me joining the project of Lars and Sung-Soo. I then came to experience Lars' mode of working in practice. We spent quite a few sessions in front of his whiteboard that fall term. My experience with higher spin solved another problem they had. Comparing to the covariantly known gravity counter-terms, there should be just two such cubic terms at derivative order $4$. Lars and Sung-Soo found extra terms parameterized by different structure of $\partial^+$. It took quite some effort to convince Lars of their origin. Consider a spin $4$ cubic interaction. Either enhancing it to a supersymmetric $N=16$ theory, or including mixed symmetry fields on the light-front, there will be lower spin components interacting with four derivatives. In particular helicity $2$ fields. Such higher spin remnants, ``pollute'' the spin $2$ counterterm computation. 

Or perhaps, one should turn the argument around. The presence of these ``extra'' counterterms -- which from the modern point of view of \emph{Higher Spin Gravity} are not really counterterms at all, but just part of the infinite set of cubic higher spin interactions -- are responsible for the exceptionally good quantum properties of the light front theory.

But back to the story-line: having got started on the light-front again, I decided to take up a problem that had been on my mind ever since I derived the cubic covariant vertex for higher spin in the BRST formulation. Working on this over the summer of 2011, I rediscovered the minimal gravitational interactions of higher spin fields on the light-front. The issue was confounded by a comments in papers by Metsaev, claiming that such interactions does not exist in dimensions $D>4$ \cite{Metsaev2007fb,Metsaev1993a}. However, in the paper \cite{Metsaev1993a}, it says that in $D=4$, the component interactions coincide with those of our paper \cite{BBL1987}, thus implicitly implying minimal gravitational couplings. Their existence is now acknowledged.

\section*{Renaissance}\label{sec:Renaissance}

The renaissance for light-front higher spin is partly connected to The Vasiliev theory. Around the millennium it attracted the attention of younger theoreticians. The newly formulated AdS/CFT conjectures \cite{Maldacena1997a}, added to the interest, as dualities between interacting higher spin gauge fields in the bulk of AdS space-time had their CFT dual on the boundary \cite{Sundborg2001a,KlebanovPolyakov2002}. 

However, the renewed interest in higher spin theory spilled over to the Minkowski theory and to what is called the Fronsdal program. Furthermore, the no-go theorems -- so notorious for the Minkowski theory -- soon found their AdS counterparts \cite{MaldacenaZhiboedov2013,BekaertBoulangerLeclercq2010a,RoibanTseytlin2017a,BoulangerSundellLeclerc2008a,SleightTaronna2018a}. The special status of the AdS theory was replaced by a more balanced view of the problems and prospects of higher spin theory. Even the positive light-front results had their AdS counterparts \cite{MetsaevLFAdS2018}. In particular, the  non-locality problems at the quartic level are present in all formulations, and the AdS theory seems to have problems already at the cubic level. The AdS non-locality problems are contentious at the time of writing, and it is best to let time pass before a verdict. However, it seems to be well recognized by now that AdS is not very special for higher spin gauge theory. More special is the choice of field variables; Fronsdal covariant tensors, Penrose twistor-like two-component fields or Dirac light-front fields.
 
It was only a matter of time before the searchlight was aimed at the light-front theory. One reason was the realization -- long overdue -- that there existed a cubic closed local massless higher spin theory with interactions, including gravitational. To start the story of the light-front renaissance, we have to go back to the two papers by R. Metsaev -- already mentioned --  on quartic interactions \cite{MetsaevQuartic1,MetsaevQuartic2} in 1991. The first paper concerns the quartic analysis for fields of even spin. The paper is sketchy, and although it defines the formalism used, it provides few details on the actual computations. The ``answer'' is not explicitly stated. And since we did not pay the paper the deserved attention, there was no-one else around to dig out the ``answer'' at the beginning of the dark ages.

In general terms -- and this is true for all perturbative Noether procedures to find interactions whatever formalism used -- computing at the second order in interaction one finds the following: the commutator of two cubic interactions (which can be computed since the cubics are supposed to be known up to relative coefficients) must be balanced by the commutator of the free kinetic term and the unknown quartic interactions. Heuristically 
\begin{equation}\label{eq:QuarticEquation}
[\mathtt{free},\mathtt{quartic}]=-[\mathtt{cubic},\mathtt{cubic}]
\end{equation}

The question is whether this equation can be solved without brutally inverting $\mathtt{free}$ producing a severe non-locality $1/\mathtt{free}$. Actually, there are two equations to solve on the light-front, one involving the Hamiltonian $H$ and the dynamical Lorentz generator $J^-$ (or the non-independent complex conjugated equation), and one involving the two dynamical Lorentz generators $J^-$ and $\bar{J}^-$. 

As for the crucial contents of the first Metsaev quartic paper, an explicit formula for the commutator $[J^-,H]_4=0$ at the quartic order is given (Formula (27) in Section 3). After this, I find the paper hard to understand. The formulas are only sketched and the logic is not very clear, and I will not try to disentangle it here. On the face of it, the paper says that ``something'' is possible only if the coefficients $C^{\lambda_1,\lambda_2,\lambda_3}$ for the $\lambda_1$-$\lambda_2$-$\lambda_3$ interactions are chosen in a particular way
\begin{equation}\label{eq:MetsaevCoefficients}
C^{\lambda_1,\lambda_2,\lambda_3}=\frac{l_p^{\lambda_1+\lambda_2+\lambda_3-1}}{\Gamma(\lambda_1+\lambda_2+\lambda_3)}
\end{equation}
with $l_p$ a constant of dimension of length.  This ``something'' can be read out as that all terms in $[\mathtt{cubic},\mathtt{cubic}]$ cancel. This means that the theory is cubic, but the paper does not say so. Instead it goes on to write a formal solution to equation \eqref{eq:QuarticEquation}, i.e. $[J^-,H]_4=0$, by dividing by $\mathtt{free}$. With all due respect, this part of the paper is not clear; at least I couldn't understand it when I read it back in 1991. The short Section 4 of the paper (Discussion) adds to the confusion by essentially writing the quartic contributions to the dynamical Lorentz generators in terms of quadratic factors of the coefficients \eqref{eq:MetsaevCoefficients}. However, shouldn't these terms cancel or add up to zero for this particular choice of coefficient? But I may be missing the point.

The second paper treats also odd spin fields by introducing internal indices. It provides some more details on the quartic computations, and contains a discussion of whether the quartic non-locality can be avoided, or not. Rather than reviewing the paper, we will rephrase the discussion with the help of the ``renaissance'' paper of D. Ponomarev and E. Skvortsov. 

The  Ponomarev-Skvortsov paper \cite{PonomarevSkvortsov2016a} clarifies the situation and digs out what was only implicit in the Metsaev papers. It is in this paper that it is explicitly realized and stated that there exist a purely cubic massless higher spin theory on the light-front provided that the the cubic coefficients are chosen according to formula \eqref{eq:MetsaevCoefficients}. The theory is chiral in the sense that only half of the interaction terms of formulas \eqref{eq:CubicInteractionLightFront} and \eqref{eq:IngemarFormulae} are retained, dropping the c.c. terms. Both helicities are present, but in an asymmetric way.

For the known cubic interactions, it is clear that the interactions are given by polynomials in the transverse momentum variables $\mathbb{P}$ and $\mathbb{\bar{P}}$ with rational functions of the $+$ direction momentum $\gamma$ as coefficients. If these restrictions can be sustained in higher orders of interaction is not yet known. What is clear is that already at the quartic level the (current exchange) interactions involve infinite sums of positive powers of transverse momenta \cite{MetsaevQuartic1,MetsaevQuartic2,PonomarevSkvortsov2016a}. Commuting two cubic level transformations (on the right hand side) to get the source for the quartic level (on the left hand side) it is clear that one gets contributions with an infinite number of momentum factors (see equation \eqref{eq:QuarticEquation}). This can be interpreted as some sort of weak non-locality, although each term in the expansion is local, as pointed out in \cite{PonomarevSkvortsov2016a}. The crucial question is whether the quartic level equation can be solved without bringing in inverse powers of transverse momenta coming from inverting $\mathbb{P}\mathbb{\bar{P}}$ in the free level transformation.

In 2015, Lars invited me to a higher spin conference in Singapore hosted by the Nanyang University. I had just started to study amplitude methods and got the partly brilliant idea that the quartic vertex could be approached from the amplitude side. And of course it can, but not in the naive way that I envisaged. I gave a talk, but soon realized that I had made a fool of myself. When writing up my contribution, I corrected that. Anyway, in the discussion after my talk, V. Didenko mentioned the 1990 paper by Metsaev studying the quartic level. 

I first thought that this was the initial impetus igniting the renaissance. However, in private  discussions E. Skvortsov I understand that he and D. Ponomarev had already started to study the old light-front literature from early and late antiquity by the time of the Singapore workshop. It was not at all straightforward to dig out the existence of the cubic theory and required a complete recomputation of the whole setup. The Ponomarev-Skvortsov paper therefore reviews and reworks the old millennium light-front theory and provides useful details of computation. It contains a derivation (quite tricky and referred to an appendix) of the particular form of the coupling factors \eqref{eq:MetsaevCoefficients} that yields the cubic chiral theory. The discussion of the quartic deformation equation is made much more explicit.

\section*{Enlightenment: The cubic chiral theory}\label{subsec6:CubicChiralTheory}

The cubic chiral theory that emerges out of the deliberations above, has come to be known under the name \emph{chiral higher spin gravity}\index{chiral higher spin gravity}. Whether it will eventually turn out to be a  sub-sector of a full, unitary and perhaps local, higher spin gravity theory, or if it will remain on its own, is unknown at the time of writing. The theory has been researched along two directions: quantum properties and covariantization. Its cubic interaction is 
\begin{equation}\label{eq6:CubicChiralInteraction}
\sum_{\lambda_1,\lambda_2,\lambda_3}\int {\textstyle{\prod_nd^{3}p_n}}\delta^{3}({\textstyle{\sum_n p_n}})\frac{l_p^{\lambda_1+\lambda_2+\lambda_3-1}}{\Gamma(\lambda_1+\lambda_2+\lambda_3)}\frac{\bar{\mathbb{P}}^{\lambda_1+\lambda_2+\lambda_3}}{\gamma_1^{\lambda_1}\gamma_2^{\lambda_2}\gamma_3^{\lambda_3}}Tr\big[\Phi_{p_1}^{\lambda_1}\ldots\Phi_{p_3}^{\lambda_3}\big]
\end{equation}
where the trace allows for non-trivial matrix valued odd spin fields. 

The provenance of this simple higher spin theory is the initial discovery in the 1980's of the cubic couplings themselves \cite{BBB1983a,BBL1987}, the quartic analysis in the early 1990's that produced the particular numerical coupling constants \cite{MetsaevQuartic1,MetsaevQuartic2}, and the mid 2010's renovation \cite{PonomarevSkvortsov2016a} that elicited the picture.

\paragraph*{Quantum properties}

As for quantum properties, there is a series of papers investigating the theory. The first one is a letter \cite{SkvortsovTranTsulaiaQI}. It is a bit short on details, but it reports the following three results: 

\emph{First}, due to the particular form of the three-point couplings, all on-shell tree amplitudes vanish, and the $S$-matrix is $1$. As the authors point out, this makes the theory consistent with the no-go theorems at least at the tree level. Remember that the Weinberg low energy theorem forbids massless higher spin fields to have long-range effects, while the Coleman-Mandula theorem forbids the $S$-matrix from having symmetry generators that transform as tensors under the Lorentz group.

\emph{Second}, all vacuum diagrams (diagrams without external states) vanish. For the one-loop ``bubble'' diagram, this hinges on the need to regularize the total number $\nu$ of degrees of freedom. This sum is infinite, but may be regularized to zero \cite{BeccariaTseytlin2015a}
\begin{equation}\label{eq6:RegularizedSumOf States}
\nu=\sum_\lambda=1+2\sum_{s=1}^\infty 1=1+2\zeta(0)=0
\end{equation}
where $\zeta(x)$ is the Riemann zeta-function. Using the value $\zeta(0)=-1/2$ one may assign a value to the divergent sum $1+1+1+\ldots$ that is the naive value at $0$ of $\zeta(x)=\sum_{n=1}^\infty n^{-x}$. According to the paper, all other vacuum diagrams vanish without the need to regularize, instead relying on the particular form of the coupling factors. Such \emph{zeta-function regularization}\index{zeta-function regularization} is not uncommon in theories with infinite number of states. It has been used in string theory as well \cite{BrinkNielsen1973} -- by Lars and Holger Bech Nielsen -- where the critical dimension was computed by using the value $\zeta(-1)=-1/12$ for the infinite sum $1+2+3+\ldots$. 

\emph{Third}, the paper argues that all loop diagrams with external legs vanish given that the total number of degrees of freedom is regularized to zero.

The second paper \cite{SkvortsovTranTsulaiaQII} and third paper \cite{SkvortsovTranQIII}, provide much more detail and strengthens and extends the result the first paper. It thus seems that the cubic chiral theory provides us with an UV-finite higher spin theory with a trivial $S$-matrix consistent with the Minkowski no-go theorems. All the same, there are actually interactions, gravitational as well, albeit cubic and non-unitary. It is still to early to know if this is just an interesting curiosity, or if the theory is part of a larger higher spin quantum theory of gravity, or if the theory serves some other unknown function in fundamental physics.

\paragraph*{Covariantization}

There is ongoing work to covariantize the cubic chiral theory \cite{KrasnovSkvortsovTranCovI,SkvortsovVanDongenCovIIa,SharapovSkvortsovSukhanovVanDongenCovIIb}. It should not be expected that this can be done using ordinary Lorentz tensors for the covariant higher spin gauge fields, as the no-go results are numerous and well-established. Instead the procedure is based on twistor theory, or rather, on two-component spinor language as introduced in the book \cite{PenroseRindlerVol1} and developed by several authors (for references, see the papers cited above). It would take us to far to enter into this fascinating subject, suffice it to note a crucial basic idea without writing any formulas. 

As was discussed in great detail a long time by Weinberg \cite{WeinbergS1964b} there are several ways to describe massless  arbitrary spin particle in field theory, corresponding to which representation of the Lorentz group one employs. A simple example is spin $1$ which may be described by the gauge invariant field strength $F_{\mu\nu}$, or by the the gauge potential $A_\mu$. In practice, both are used and they are related in the well-known way. However, one may imagine putting away this relationship (for a while) and see how far one gets by either picture. More to the point, one may imagine describing the two helicities in different ways: one helicity by the field-strength, the other with the gauge potential. This is actually quite easy to do in the two-component spinor formulation, and has been employed in the twistor approach to field theory. Thus one may say that in covariantizing the light-front cubic chiral theory, a way forward would be to retain physical fields (field strengths) for half the helicities, while introducing gauge fields (potentials) for the other half of the helicities. Indeed, in this way the theory becomes chiral.  

The actual implementation of these ideas runs parallel to self-dual Yang-Mills and self-dual gravity theories. However, at the time of writing, it seems that one eventually has to resort to equations of motion and the formalism of free differential algebras, not unlike the formalism of the Vasiliev theory. Thus, one may fear that the ``Pandora box'' of infinite sets of auxiliary fields is opened again. Perhaps not unexpected, certainly interesting, but in my opinion, running away from the initial simplicity of the light-front theory. However, for the cubic chiral theory, locality is under control. This research is evolving rapidly, and the end is not yet in sight.\footnote{Developments since the spring of 2023 when this text was initially prepared has not been included here.}

\section*{Never prepare -- at least not too much}\label{sec:NeverPrepare}
Every Friday there was a group meeting. Everyone took turns giving talks. It started at one end of one of the corridors in the beginning of the fall term and then went down corridor after corridor relentlessly. It was easy to figure out when your office was up. And everyone gave talks: PhD students, professors, associate professors, senior lecturers, the janitor (well not really), guests, et cetera. And as I remember it, it was all very generous -- no flaunting of smartness -- just serious physics where the focus was more on the questions than on the answers. And as Ingemar reminded me off, they lasted for two full hours without a break, from after lunch at one o' clock till coffee at three. Still people very seldom fell asleep.

One fall term Lars came back from the US with an apparently crazy idea. For the Friday group meetings we were not allowed to prepare our talks. How that was going to be checked wasn't said. But the idea was that you should know your subject so well that you should be able to talk about it without preparing. The talks were mostly white-board although it happened that transparencies were used. 

One un-happy guest did that -- used transparencies -- and in doing so underlined the basic fact that if you do prepare, you should prepare right (I don't think the not-prepare rule applied to guests though). Anyway, this particular unfortunate guest had his transparencies rolled up in a tight roll. And of course they didn't want to unroll on the overhead projector. There was nothing heavy enough in the room, still small enough not to hide the formulas, to hold the darned plastic sheets down. Those particular two hours were two long hours.

I don't know to what extent the idea of not preparing was implemented generally, but it took hold of my imagination. I must have given some Friday talk unprepared or at least not preparing all the details. The real point became clear to me later on when I started to teach more. It is indeed very natural when it comes to teaching.

A few years later when I started to teach at a gymnasium I found that I didn't have to prepare, and I didn't like to prepare. Preparing classes was boring. If I tried to sit down and prepare details I soon got bored. It was as if I needed an audience. So I stopped preparing, just contending myself with a general plan for the course and briefly thinking through what to do in class. When I returned to academia I found that this method worked well there too. I have never prepared a lecture or exercise class, just quickly made up my mind what to treat. It is better -- I don't know if Lars said this, but this is how I interpreted it -- to use your time learning new things rather than preparing yourself for talking about things you already should know. It has saved me a lot of time over the years and made it possible to do research on a teaching position.

It's time for some last words. As I wrote to Lars last summer, I am deeply grateful for the privilege of having been one of his students. Most of all, for the opportunity to study and work within the subject of theoretical physics, the intellectual territory of the popular science adventure books I read as a teenager.

\pagebreak


\begin{thebibliography}{10}

\bibitem{BrinkLindgrenNilsson1983}
L.~Brink, O.~Lindgren, and B.E.W. Nilsson.
\newblock The ultra-violet finiteness of the {N}=4 {Y}ang-{M}ills theory.
\newblock {\em Phys. Lett.}, 123B:323--328, 1983.

\bibitem{Mandelstam1983}
S.~Mandelstam.
\newblock Light-cone superspace and the ultraviolet finiteness of the {N}=4
  model.
\newblock {\em Nucl. Phys. B}, 213:149--168, 1983.

\bibitem{BrinkLindgrenNilsson1983N4}
L.~Brink, O.~Lindgren, and B.E.W. Nilsson.
\newblock {N}=4 {Y}ang-{M}ills theory on the light cone.
\newblock {\em Nucl. Phys. B}, 212:401--412, 1983.

\bibitem{BBB1983a}
A.~K.~H. Bengtsson, I.~Bengtsson, and L.~Brink.
\newblock Cubic interaction terms for arbitrary spin.
\newblock {\em Nucl. Phys. B}, 227:31--40, 1983.

\bibitem{BBB1983b}
A.~K.~H. Bengtsson, I.~Bengtsson, and L.~Brink.
\newblock Cubic interaction terms for arbitrarily extended supermultiplets.
\newblock {\em Nucl. Phys. B}, 227:41--49, 1983.

\bibitem{Dirac1949FormsRelDyn}
P.~A.~M. Dirac.
\newblock Forms of relativistic dynamics.
\newblock {\em Rev. Mod. Phys.}, 21:392--399, 1949.

\bibitem{Dirac1933a}
P.A.M. Dirac.
\newblock The {L}agrangian in quantum mechanics.
\newblock {\em Physikalische Zeitschrift der Sowjetunion}, Band 3, Heft
  1:64--72, 1933.

\bibitem{Feynman1948PathInt}
R.P. Feynman.
\newblock Space-time approach to non-relativistic quantum mechanics.
\newblock {\em Rev. Mod. Phys.}, 20:367--387, 1948.

\bibitem{BBL1987}
A.~K.~H. Bengtsson, I.~Bengtsson, and N.~Linden.
\newblock Interacting higher-spin gauge fields on the light front.
\newblock {\em Class. Quant. Grav.}, 4:1333--1345, 1987.

\bibitem{KohOuvry1986}
I.~G. Koh and S.~Ouvry.
\newblock Interacting gauge fields of any spin and symmetry.
\newblock {\em Phys. Lett. B}, 179:115--118, 1986.

\bibitem{AragoneDeser1971}
C.~Aragone and S.~Deser.
\newblock Constraints on gravitationally coupled tensor fields.
\newblock {\em Nuovo Cimento}, 3A:709, 1971.

\bibitem{AragoneDeser1980a}
C.~Aragone and S.~Deser.
\newblock Consistency problems of spin-2-gravity coupling.
\newblock {\em Nuovo Cimento}, 57B:33, 1980.

\bibitem{AragoneDeser1979a}
C.~Aragone and S.~Deser.
\newblock Consistency problems of hypergravity.
\newblock {\em Phys. Lett. B}, 86:161--163, 1979.

\bibitem{AragoneDeser1980b}
C.~Aragone and S.~Deser.
\newblock Higher spin vierbein gauge fermions and hypergravities.
\newblock {\em Nucl. Phys. B}, 170:329--352, 1980.

\bibitem{AKHB1988}
A.~K.~H. Bengtsson.
\newblock {BRST} approach to interacting higher-spin gauge fields.
\newblock {\em Class. Quant. Grav.}, 5:437--451, 1988.

\bibitem{LarsBrink1987HigherSpin}
L.~Brink.
\newblock Field theories for higher spin.
\newblock In I.~A. Batalin, C.~J. Isham, and G.~A. Vilkovisky, editors, {\em
  Quantum Field Theory and Quantum Statistics}, volume~2, pages 197--207. Adam
  Hilger, Bristol, 1987.
\newblock Essays in Honour of the Sixtieth Birthday of E. S. Fradkin.

\bibitem{AKHB1985}
A.~K.~H. Bengtsson.
\newblock Gauge invariance for spin-3 fields.
\newblock {\em Phys. Rev. D}, 32:2031--2036, 1985.

\bibitem{Fronsdal1979conf}
C.~Fronsdal.
\newblock Some open problems with higher spins.
\newblock In P.~van Nieuwenhuizen and D.~Z. Freedman, editors, {\em
  Supergravity}, pages 245--249. North-Holland Publishing Company, 1979.

\bibitem{OuvryStern1986a}
S.~Ouvry and J.~Stern.
\newblock Gauge fields of any spin and symmetry.
\newblock {\em Phys. Lett. B}, 177:335--340, 1986.

\bibitem{AKHB1986a}
A.~K.~H. Bengtsson.
\newblock A unified action for higher spin gauge bosons from covariant string
  theory.
\newblock {\em Phys. Lett. B}, 182:321--325, 1986.

\bibitem{Siegel1986a}
W.~Siegel.
\newblock Classical superstring mechanics.
\newblock {\em Nucl. Phys. B}, 263:93--104, 1986.

\bibitem{Metsaev2007fb}
R.~R. Metsaev.
\newblock Cubic interaction vertices for fermionic and bosonic arbitrary spin
  fields.
\newblock {\em Nucl. Phys. B}, 859:13--69, 2012.
\newblock arXiv:0712.3526.

\bibitem{Metsaev1993a}
R.~R. Metsaev.
\newblock Generating function for cubic interaction vertices of higher spin
  fields in any dimension.
\newblock {\em Modern Physics Letters A}, 8(25):2413--2426, 1993.

\bibitem{Maldacena1997a}
J.~Maldacena.
\newblock The large {N} limit of superconformal field theories and
  supergravity.
\newblock {\em Adv. Theor. Math. Phys.}, 2:231252, 1998.
\newblock arXiv:hep-th/9711200.

\bibitem{Sundborg2001a}
B.~Sundborg.
\newblock Stringy gravity, interacting tensionless strings and massless higher
  spins.
\newblock {\em Nucl. Phys. Proc. Suppl.}, 102:113--119, 2001.
\newblock arXiv:hep-th/0103247.

\bibitem{KlebanovPolyakov2002}
I.~R. Klebanov and A.~M. Polyakov.
\newblock {AdS} dual of the critical $o(n)$ vector model.
\newblock {\em Phys. Lett. B}, 550:213--219, 2002.
\newblock arXiv:hep-th/0210114.

\bibitem{MaldacenaZhiboedov2013}
J.~Maldacena and A.~Zhiboedov.
\newblock Constraining conformal field theories with a slightly broken higher
  spin symmetry.
\newblock {\em Class. Quant. Grav.}, 30:104003, 2013.
\newblock arXiv:1204.3882.

\bibitem{BekaertBoulangerLeclercq2010a}
X.~Bekaert, N.~Boulanger, and S.~Leclercq.
\newblock Strong obstruction of the {B}erends-{B}urgers-van {D}am spin-3
  vertex.
\newblock {\em Journal of Physics A: Mathematical and Theoretical},
  43(18):185401, 2010.
\newblock arXiv:1002.0289.

\bibitem{RoibanTseytlin2017a}
R.~Roiban and A.~A. Tseytlin.
\newblock On four-point interactions in massless higher spin theory in flat
  space.
\newblock {\em JHEP}, 04:139, 2017.
\newblock arXiv:1701.05773.

\bibitem{BoulangerSundellLeclerc2008a}
N.~Boulanger, P.~Sundell, and S.~Leclerc.
\newblock On the uniqueness of minimal coupling in higher-spin gauge theory.
\newblock {\em JHEP}, 0808:056, 2008.
\newblock arXiv:0805.2764.

\bibitem{SleightTaronna2018a}
C.~Sleight and M.~Taronna.
\newblock Higher-spin gage theories and bulk locality.
\newblock {\em Phys. Rev. Lett.}, 121:171604, 2018.
\newblock arXiv:1704.07859.

\bibitem{MetsaevLFAdS2018}
R.R. Metsaev.
\newblock Light-cone gauge cubic interaction vertices for massless fields in
  {AdS(4)}.
\newblock {\em Nucl. Phys. B}, 936:320--351, 2018.
\newblock arXiv:1807.07542.

\bibitem{MetsaevQuartic1}
R.~R. Metsaev.
\newblock Poincar{\'e} invariant dynamics of massless higher spins: Fourth
  order analysis on mass shell.
\newblock {\em Mod. Phys. Lett.}, A6:359--367, 1991.

\bibitem{MetsaevQuartic2}
R.~R. Metsaev.
\newblock S-matrix approach to massless higher spins theory: {II}. the case of
  internal symmetry.
\newblock {\em Mod. Phys. Lett.}, A6:2411--2421, 1991.

\bibitem{PonomarevSkvortsov2016a}
D.~Ponomarev and E.~Skvortsov.
\newblock Light-front higher-spin theories in flat space.
\newblock {\em Journal of Physics A: Mathematical and Theoretical}, 50:095401,
  09 2016.
\newblock arXiv:1609.04655.

\bibitem{SkvortsovTranTsulaiaQI}
E.~Skvortsov, T.~Tran, and M.~Tsulaia.
\newblock Quantum chiral higher spin gravity.
\newblock {\em Phys. Rev. Lett.}, 121:031601, 2018.
\newblock arXiv:1805.00048.

\bibitem{BeccariaTseytlin2015a}
M.~Beccaria and A.~A. Tseytlin.
\newblock On higher partition functions.
\newblock {\em J. Phys. A: Mathematical and Theoretical}, 48, 2015.
\newblock arXiv:1503.08143.

\bibitem{BrinkNielsen1973}
L.~Brink and H.~B. Nielsen.
\newblock A simple physical interpretation of the critical dimension of
  space-time in dual models.
\newblock {\em Phys. Lett. B}, 45:332--336, 1973.

\bibitem{SkvortsovTranTsulaiaQII}
E.~Skvortsov, T.~Tran, and M.~Tsulaia.
\newblock More on quantum chiral higher spin gravity.
\newblock {\em Phys. Rev. D}, 101:106001, 2020.
\newblock arXiv:2002.08487.

\bibitem{SkvortsovTranQIII}
E.~Skvortsov and T.~Tran.
\newblock One-loop finiteness of chiral higher spin gravity.
\newblock {\em JHEP}, 07:021, 2020.
\newblock arXiv:2004.10797.

\bibitem{KrasnovSkvortsovTranCovI}
K.~Krasnov, E.~Skvortsov, and T.~Tran.
\newblock Actions for self-dual higher spin gravities.
\newblock {\em JHEP}, 08:076, 2021.
\newblock arXiv:2105.12782.

\bibitem{SkvortsovVanDongenCovIIa}
E.~Skvortsov and R.~Van Dongen.
\newblock Minimal model of field theories: Chiral higher spin gravity.
\newblock {\em Phys. Rev. D}, 106:045006, 2022.
\newblock arXiv:2204.10285.

\bibitem{SharapovSkvortsovSukhanovVanDongenCovIIb}
A.~Sharapov, E.~Skvortsov, A.~Sukhanov, and R.~Van Dongen.
\newblock Minimal model of chiral higher spin gravity.
\newblock {\em JHEP}, 09:134, 2022.
\newblock arXiv:2205.07794.

\bibitem{PenroseRindlerVol1}
R.~Penrose and W.~Rindler.
\newblock {\em Spinors and space-time, volume 1}.
\newblock Cambridge Univ. Press, 1984.

\bibitem{WeinbergS1964b}
S.~Weinberg.
\newblock Feynman rules for any spin. {II}. {M}assless particles.
\newblock {\em Phys. Rev.}, 134:B882--B896, 1964.

\end{thebibliography}

\end{document}